\documentclass[runningheads,a4paper]{llncs}

\usepackage{amssymb}
\usepackage{graphicx}

\usepackage{url}
\urldef{\mailsa}\path|{asealfon,katesot}@mit.edu|   
\newcommand{\keywords}[1]{\par\addvspace\baselineskip
\noindent\keywordname\enspace\ignorespaces#1}

\begin{document}

\mainmatter  
\title{Agreement in Partitioned Dynamic Networks}

\titlerunning{Agreement in Partitioned Dynamic Networks}

\author{Adam Sealfon
\and Aikaterini Sotiraki}

\authorrunning{Agreement in Partitioned Dynamic Networks}
\institute{Massachusetts Institute of Technology\\
\mailsa\\}

\maketitle

\begin{abstract}
In the dynamic network model, the communication graph is assumed to be connected in
every round but is otherwise arbitrary.  We consider the related setting of {\em $p$-partitioned
dynamic networks}, in which the communication graph in each round consists of at most $p$
connected components. We explore the problem of $k$-agreement in this model for $k\geq p$. 
We show that if the number of processes is unknown then 
 it is impossible to achieve $k$-agreement
for any $k$ and any $p\geq 2$. Given an upper bound $n$ on the number of processes, we provide
algorithms achieving
$k$-agreement in $p(n-p)$ rounds for $k=p$ and in $O(n/\epsilon)$ rounds
for $k=\lceil (1+\epsilon)p \rceil$.
\end{abstract}

\section{Introduction}
\label{intro-sec}

Dynamic graphs are a model for distributed algorithms which were introduced by Kuhn, Lynch and Oshman \cite{KLO10}. 
In this paper we explore the capabilities and limitations of
a modification to the dynamic graph model addressing additional challenges arising in wireless communication.

In the dynamic graph model, the network is assumed merely to be connected in each round,
with no additional assumptions about consistency from round to round. We weaken this
assumption further, allowing the network to consist of more than one connected component.  Formally,
\begin{definition}
A dynamic graph $G=(V,E)$ is said to be {\em $p$-partitioned} if at each round $t$, it consists of at most $p$ connected components.
\end{definition} 
\noindent Processes communicate in synchronous rounds using local broadcast. The edges in each round are chosen by an adaptive adversary.

In this setting, many of the problems usually considered in the dynamic network model cannot be solved. 
In particular, tasks such as token dissemination,  leader election and consensus which cannot be solved in partitioned static networks clearly are also impossible in partitioned dynamic networks. 
We consider the problem of $k$-agreement for constant $k$, 
which can be solved in static $p$-partitioned networks as long as $k\geq p$.
The conditions for $k$-agreement are the following:
\begin{enumerate}
\item Agreement: All decision values are in $W,$ where $W$ is a subset of the initial values with $|W|= k$.
\item Validity: Any decision value is the initial value of some process.
\item Termination: All processes eventually decide.
\end{enumerate} 

We show that $k$-agreement is not possible in the setting of $p$-partitioned dynamic networks if the 
number of processes is unknown, but that it can be achieved 
for any $k\geq p$ given an upper bound on 
the number of processes.
Our results are qualitatively different from the case of ordinary dynamic networks, for
which there are known consensus protocols which do not assume knowledge of the size of the network \cite{O12}.

\begin{theorem}
\label{thm:impossibility}
For all $p\geq 2$, $k\geq 1$ there is no algorithm which will solve $k$-agreement
on $p$-partitioned dynamic graphs given no information about the size of the network.
\end{theorem}

\begin{theorem}
For any $p\geq 1$, we can 
solve $p$-agreement in $p(n-p-1)+1$ rounds on any $p$-partitioned dynamic graph,
where $n$ is a known upper bound on the number of vertices.
\end{theorem}

\begin{theorem}
For any $\epsilon>0$, $p\geq 1$, we can solve
$\lceil (1+\epsilon) p \rceil$-agreement in $O(n/\epsilon)$ rounds on any $p$-partitioned
dynamic graph,
where $n$ is a known upper bound on the number of vertices.
\end{theorem}

\section{Impossibility with unknown network size}
\label{partitioned-impossibility-subsec}
It suffices to consider the case $p=2$. Let $k\geq 1$, and assume
there exists some algorithm which solves $k$-agreement on $2$-partitioned dynamic networks.
We construct an execution with $k+1$ distinct decision values 
$1,\ldots,k+1$. 

Consider some process $v$. If after some point in the execution
$v$ has no neighbors for sufficiently many rounds, 
$v$ will eventually decide on some value it has heard and terminate. If all values it has received up to this point are equal to its input value, then by the validity requirement this must be its decision value.

We divide the execution into $k+1$ phases. In phase $i$, the communication graph $G$ 
consists of a single isolated process $a_i$ and a path $P_i$ connecting all the other processes.
The communication graph will stay unchanged within each phase,
and at the end of phase $i$ process $a_i$ will decide on the value $i$. 

Let $t>0$ be sufficiently large, as determined later.
Path $P$ consists of $k+1$ segments, 
each containing $2t+1$ processes.
Each process in the $i$th segment has input value $i$. 
Let $a_i$ be the middle vertex in segment $i$, and let $a_i', a_i''$ be its neighbors in $P$. 
In phase $i$, $G$ will be produced from $P$ by 
turning $a_i$ into an isolated vertex and adding an edge 
connecting $a_i'$ and $a_i''$ to form path $P_i$. 

If phase $i$ begins on round $t_i<t$, then at the beginning of the phase
process $a_i$ has heard from $2t_i$ other processes, all of which share
its input value $i$. During phase $i$ process $a_i$ has no neighbors, 
so as observed above, it must decide on value $i$ within finitely many
rounds. 
Note that this number of rounds is independent of $t$ as long as $t>t_i$. 
Consequently, taking $t>t_{k+1}$, we have produced an execution in which 
process $a_i$ decides on value $i$ for each $i$, proving the impossibility result.

\section{Algorithms for known network size}
\label{partitioned-algorithm-subsec}
The proof in the previous setting relied heavily on processes not knowing the total number $n$ of processes.
This is no coincidence. In this section we show that if processes do know the number of process $n$ or
an upper bound on $n$, they can solve $k$-agreement in a $p$-partitioned dynamic network for any $k\geq p$. 

The algorithm is extremely simple. Each process $v$ stores the smallest value $m_v$
it has seen so far. On each round, the process
broadcasts this value to its neighbors. 
If it receives a smaller value, it updates $m_v$. 
After some fixed number $\gamma$ of rounds,
each process outputs the smallest value it has seen.

We now show that with $\gamma=p(n-p-1)+1$ rounds, this algorithm
solves $p$-agreement on $p$-partitioned graphs with at most $n$ vertices.
Since the number of rounds depends on $n$, this does not violate the 
impossibility result of the previous section. 
Consider some dynamic network $G=(V,E)$ with initial inputs $\vec{x}$.
Let $m_v(t)$ be the value of variable $m_v$ at the beginning of round $t$,
and let $S(t)=\{m_v(t):v\in V\}$ be the set of the minimum values heard by each process
by some round $t$. 
For each $i\leq p$, let $s_i(t)$ be the $i$th smallest value in $S(t)$, let 
$V_i(t) \subseteq V$ be the set of processes with value $s_i(t)$ in round $t$,
and let $A_i(t) = |V_1(t)|+\cdots+|V_i(t)|$ be the number of processes with value at most $s_i(t)$.
Let $V_{p+1} (t) = V\setminus (V_1(t)\cup\cdots\cup V_p(t))$.
Observe that the value $s_i(t)$ depends on the round,
so the value of the processes in the set $V_i(t)$ may be different in different rounds.
We make use of the potential function
$$\Phi(t) = p\cdot |V_1(t)| + (p-1)\cdot |V_2(t)| + \cdots + 2\cdot |V_{p-1}(t)| + |V_p(t)| = A_1(t)+\cdots+A_p(t).$$
We will show that as long as
there are more than $p$ values in $S(t)$, $\Phi(t+1) > \Phi(t)$.

Note that $S(t+1)\subseteq S(t)$ and that for any process $v$ the variable
$m_v$ is nonincreasing. Consequently for all $i$ we have that the variable 
$A_i(t)$ never decreases.
It remains to show that as long as $|S(t)|>p$,
there is an $i$ such that $A_i(t+1) > A_i(t)$.  
But the communication graph in round $t$ consists of at most $p$ connected components,
so there must be a pair of vertices $u\in V_j(t), v\in V_\ell(t)$ with $j<\ell$ and $j\leq p$
such that $uv \in E(t)$. Then in round $t+1$ process $v$ has value at most $s_j(t)$,
so $A_j(t+1) > A_j(t)$. 
Since each $A_i(t)$ is nondecreasing,
this implies that $\Phi(t+1) > \Phi(t)$. 

But the initial value $\Phi(0) \geq p + (p-1) + \cdots + 1 = p(p+1)/2$.
The maximum value of $\Phi$ in any round $t'$ for which $|S(t')|>p$ occurs for 
$|V_1(t')|=n-p$, $|V_2(t')|=|V_3(t')|=\cdots=|V_p(t')| = 1$, so if $|S(t)|>p$ then
we have that
$$\Phi(t) \leq p(n-p) + p(p-1)/2.$$
Since $\Phi(t)$ must increase by at least one in each round as long as 
$|S(t)|>p$, after at most
$$\gamma = 1 + \Phi(t') - \Phi(0) \leq p(n-p-1)+1$$
rounds we will have that $|S(\gamma)|\leq p$, and so the algorithm above 
achieves $p$-agreement.

We now show that if $k \geq (1+\epsilon)p$, the same algorithm
achieves $k$-agreement after $\gamma = O(n/\epsilon)$ rounds. 
Similarly to the above, 
for $i\leq k$ let $V_i(t)$ be the set of processes with the $i$th smallest value $s_i(t)$ in round $t$, and
let $V_{k+1}(t) = V \setminus (V_1(t)\cup\cdots\cup V_k(t))$.
We will show that the potential function
$$\Phi(t) = k\cdot |V_1(t)| + (k-1)\cdot |V_2(t)| + \cdots + 2\cdot |V_{k-1}(t)| + |V_k(t)|$$
increases by at least $1+k-p$ in each round.
Define a graph $G'(t)$ on vertex set
$\{V_1(t),\ldots,V_{k+1}(t)\}$ with an edge between 
$V_i(t),V_j(t)\in V'$ if there are vertices $v_i\in V_i(t)$, $v_j\in V_j(t)$ which are
connected by an edge on round $t$ in the original graph $G$. 
Since $G$ consists of at most $p$ connected components,
by construction $G'(t)$ must also consist of at most $p$
connected components. 
If $V_i(t)$ and $V_j(t)$ are in the same
connected component of $G'(t)$ for $i<j$, then we can consider
some simple path connecting them. Let the indices of the vertices on
this path be $(i=a(0), a(1),\ldots,a(r)=j)$. 
For each edge $V_{a(q)}(t)V_{a(q+1)}(t)$ in the path with $a(q+1)>a(q)$, 
we have that the potential function $\Phi$ increases by $a(q+1)-a(q)$.
Consequently a component containing $V_i(t)$ and $V_j(t)$ for $i<j$ corresponds 
to an increase in $\Phi$ by at least $j-i$. 
Therefore a connected component in $G'(t)$ consisting of at least
$s$ sets $V_i(t)$ corresponds to an increase in $\Phi$ by at least
$s-1$. There are $k+1$ sets $V_1(t),\ldots,V_{k+1}(t)$ but only $p$ connected components,
so we must have that $\Phi$ increases in each round by at least 
$k+1-p > 1+k\epsilon/(1+\epsilon)$.
But identically to the above we have that 
$\Phi(t')-\Phi(0) \leq k(n-k-1)$ in any round $t'$ for which $k$-agreement
has not yet been achieved. 
Consequently after at most
$$\gamma = 1 + \frac{k(n-k-1)}{1+k\epsilon/(1+\epsilon)} 
< 1 + \frac{(1+\epsilon)n}{\epsilon}
= O(n/\epsilon)$$
rounds the algorithm above will achieve $k$-agreement.

\section{Discussion}
We have introduced a variant of dynamic networks in which the 
network is no longer assumed to be connected in each round. 
We show that it remains possible to solve nontrivial problems
under the weaker assumption that the network 
at each round consists of at most $p$ connected components. 
In particular, we show that given an upper bound on the size of 
the network, it is possible to solve $k$-agreement in a $p$-partitioned 
dynamic network.

It would be interesting to consider whether it is possible to achieve
agreement in fewer rounds in a $p$-partitioned dynamic network. Our algorithms
solve $\lceil(1+\epsilon)p\rceil$-agreement in 
$O(n/\epsilon)$ rounds and $p$-agreement in $p(n-p-1)+1$ rounds. 
It is unclear whether this
dependence on $p$ is intrinsic or whether 
$p$-agreement can be achieved in $O(n)$ rounds regardless of $p$. 
It would also be interesting to explore whether $p$-agreement
can be achieved in fewer rounds with high probability against a nonadaptive adversary. 
It also remains open what additional problems can be solved in this model.

\subsubsection*{Acknowledgments.} We would like to thank Mohsen
Ghaffari and Nancy Lynch for helpful discussions.
This material is based upon work supported in part
by the National Science Foundation Graduate Research Fellowship under Grant No. 1122374.

\end{document}